\documentclass[10pt,preprint]{aastex}
\usepackage{apjfonts,emulateapj5}

\newcommand{\nsfull}{RX~J1856.5$-$3754}
\newcommand{\ns}{RXJ1856}
\newcommand{\Msun}{\ifmmode\mbox{M}_{\odot}\else$\mbox{M}_{\odot}$\fi}
\newcommand{\degrees}{\ifmmode^{\circ}\else$^{\circ}$\fi}
\newcommand{\amin}{\ifmmode^{\prime}\else$^{\prime}$\fi}
\newcommand{\asec}{\ifmmode^{\prime\prime}\else$^{\prime\prime}$\fi}

\slugcomment{Accepted by ApJ Letters 2002 April 1}

\shortauthors{Ransom et al.}
\shorttitle{A Deep Search for Pulsations from \nsfull}

\begin{document}

\title{A Deep Search for Pulsations from the Nearby Isolated Neutron
  Star \nsfull}

\author{Scott~M.~Ransom\altaffilmark{1, 2}, 
  Bryan~M.~Gaensler\altaffilmark{3}, and
  Patrick~O.~Slane \altaffilmark{3}}


\altaffiltext{1}{Department of Physics, McGill University, Montreal,
  QC. H3A~2T8 Canada; ransom@physics.mcgill.ca}
\altaffiltext{2}{Center for Space Research, Massachusetts Institute of
  Technology, Cambridge, MA 02139}
\altaffiltext{3}{Harvard-Smithsonian Center for Astrophysics, 
  60 Garden St., Cambridge, MA 02138; 
  bgaensler@cfa.harvard.edu, slane@head-cfa.harvard.edu}

\begin{abstract}
  We present the results of a deep search for pulsations from the
  nearby isolated neutron star \nsfull\ using the 450\,ks Director's
  Discretionary Time \emph{Chandra}\ observation completed on 2001 Oct
  15.  No pulsations were detected.  We find a 99\% confidence upper
  limit on the pulsed fraction of $\sim4.5\%$ for worst-case
  sinusoidal pulsations with frequency $\la 50$\,Hz and frequency
  derivatives $-5\times10^{-10} \le \dot{f} \le 0$\,Hz\,s$^{-1}$.  The
  non-detection of pulsations is most likely due to an unfavorable
  viewing angle or emitting geometry.  Such emitting geometries are
  much more likely to occur for more ``compact'' neutron stars which
  show increased gravitational light-bending effects.  In this case,
  the non-detection implies a radius/mass ratio for \nsfull\ of $R/M
  \la 10$\,km\,$\Msun^{-1}$.
\end{abstract}


\section{Introduction}
\label{sec:intro}

The \emph{ROSAT}\ observatory provided us with at least seven soft
X-ray emitting point sources which are likely to be radio-quiet
thermally-emitting isolated neutron stars \citep[NSs; for reviews
see][]{tpc+01, mot02}.  The nearest of these systems is \nsfull\ 
(hereafter \ns), which was first observed as an unidentified point
source in the \emph{EINSTEIN}\ Slew Survey \citep{eps+92} and later
identified as an isolated NS by \citet*{wwn96}.  Subsequent
observations in the optical and ultraviolet (UV) have strengthened
this claim by detecting a faint blue optical counterpart \citep{wm97},
measuring its proper motion \citep{wal01}, and identifying a bow-shock
$H_\alpha$ nebula around the source \citep{vk01}.  \citet{wal01} also
used \emph{Hubble Space Telescope (HST)}\ data to measure a parallax
distance of $61^{+9}_{-8}$\,pc to \ns.  However, \citet*{kva02} have
recently re-analyzed the same data and determined a distance of
$140\pm40$\,pc --- the reasons for the large discrepancy are, as of
yet, unclear.

Despite these measurements, the nature of \ns\ remains a mystery.  Is
it young or old? Is it an ordinary radio pulsar beaming away from us,
a low magnetic field neutron star accreting from the interstellar
medium, or a high-field ``magnetar''? A key to answering these
questions lies in detecting pulsations from this source, from which we
could then determine its spin-period, age, braking torque, and
magnetic field strength.  If an isolated neutron star is sufficiently
magnetized, the resulting anisotropies in the temperature distribution
of the crust can produce hotspots of X-ray emission on the surface of
the star.  If the magnetic axis is mis-aligned with the rotation axis,
these hotspots may produce a periodic modulation of the soft X-ray
intensity at the rotation period of the NS.  Even a non-detection of
these pulsations can provide useful constraints on the mass, radius,
and emitting geometry of the NS.

Searches for these pulsations were carried out by \citet{bzn+01} using
a $\sim56$\,ks \emph{Chandra X-Ray Observatory}\ observation of \ns\ 
taken on 2000 March 10 using the Low Energy Transmission Grating
Spectograph \citep[LETGS\footnote{The LETGS is composed of the High
  Resolution Camera Spectroscopy \citep[HRC-S, ][]{mck+98} and the Low
  Energy Transmission Grating (LETG).}, ][]{bgk+00}.  \citet{bzn+01}
found no pulsations and placed an upper limit on the pulsed fraction
(for frequencies $\la40$\,Hz) of $\sim8$\%.  Similar searches have
been undertaken with archival \emph{ROSAT}\ and \emph{ASCA}\ data by
\citet{pwl+02} who placed a 50\% confidence limit on the pulsed
fraction of $\sim6$\%.  It should be noted that three other radio
quiet isolated NS candidates have been reported to show X-ray
pulsations: RX~J0720.4$-$3125 with a period of 8.4\,s and a pulsed
fraction of $\sim12$\% \citep{hmb+97}, RX~J0420.0$-$5022 with a period
of 22.7\,s and a pulsed fraction of $\sim40$\% \citep{hpm99}, and
RBS~1223 with a period of 5.16\,s and a pulsed fraction of $\sim20$\%
\citep{hhs+01}.

In 2001 October, \ns\ was observed with the \emph{Chandra}\ LETGS for
an additional 450\,ks of Director's Discretionary Time (DDT) over the
course of $\sim7.25$\,days.  The purpose of the observation was to
search for features in the spectrum --- in particular, due to heavy
elements in the NS atmosphere --- in order to constrain the NS
equation of state (EOS).  However, these data are of sufficient time
resolution to allow a deep search for pulsations as well.  The data
were made public soon after the observation was completed.  We
conducted brute-force period folding and advanced Fourier analyses of
this data in order to try to identify the spin parameters of the
neutron star.  In addition, we have analyzed archival observations of
\ns\ from both \emph{ROSAT}\ and \emph{Chandra}\ in order to specify a
consistent set of upper limits to the pulsed fraction.

\section{Observations and Data Preparation}
\label{sec:prep}

The DDT observation comprised ObsIds 3382, 3380, 3381, and 3399 in
chronological order beginning on 2001 October 8 and ending on 2001
October 15 (see Table~\ref{table1}) and were taken in the standard
LETGS configuration.  The total exposure time was 450\,ks with
significant gaps between the various ObsIds giving a total duration of
$\sim 626$\,ks.

The data were prepared in a very simple manner.  We extracted all
events from within $\sim2\asec$ of the centroid of the zero-order
image from each ObsId.  The arrival times of these events were
transformed to the Solar System Barycenter using the standard
\emph{Chandra Interactive Analysis of Observations (CIAO, v2.2)}\ tool
{\tt axBary}, the nominal position of \ns\ from \citet{wal01} of ${\rm
  RA (J}2000{\rm )} = 18^{\rm h}$ 56$^{\rm m}$ 35\fs5 and ${\rm DEC
  (J}2000{\rm )} = -37\degrees\;54\amin\;36\farcs8$, and the
preliminary Level 2 orbital ephemeris file distributed with the data
release.  We binned the 90134 resulting events into 1.5\,ms time bins
to create a 420 million point time series that was used in the
coherent Fourier analysis described in \S\ref{sec:FFT}.

A wiring error in the HRC causes each event to be tagged with the
arrival time of the previous event.  Unfortunately not every event is
telemetered to the ground, so the arrival times of recorded events are
typically in error by a few ms\footnote{See {\tt
    http://asc.harvard.edu/cal/Links/Hrc/CIP/timing.html}}.  In order
to improve the accuracy of these arrival times and to restrict the
number of counts so as to make a brute-force period folding search
computationally feasible, we also reprocessed the data using the time
filtering scheme described by \citet{tbj+01}.  By shifting all arrival
times in the Level 1 event files back to the previous event, filtering
the data spatially, barycentering, and then keeping only those events
that arrived within 1\,ms of the previous event, we are guaranteed to
have all arrival times accurate to $<1$\,ms at the cost of a loss of
$\sim 95$\% of the events that were used in the Fourier analysis (see
Table~\ref{table1}).  These time filtered events were used in the
folding search described in \S\ref{sec:GLfold}.

In order to verify preparation methods for the data, we processed
events from the LETGS observation of the Crab pulsar (ObsId~759) in
the same manner as described above.  The Crab pulsar was easily
detected in the data at the expected barycentric rotation period and
with the expected pulse profile given the HRC wiring error.  In
addition, significant detection of all spin harmonics up to at least
the eighth ($\sim239$\,Hz), indicate that fast periodicities
($\gtrsim100$\,Hz) are detectable --- albeit drastically suppressed ---
in unfiltered HRC data given enough signal-to-noise.

\section{Data Analysis}
\label{sec:analysis}

In order to maximize our sensitivity to coherent pulsations with a
variety of pulse shapes, we searched the data using two very different
techniques.  For maximum sensitivity to sinusoidal pulse profiles, we
performed a coherent Fourier analysis of all the zero-order events
from the complete DDT observation.  For better sensitivity to more
complicated pulse profiles, we performed period folding searches on
the much smaller set of time-filtered events with the significance of
each trial determined by the Bayesian method developed by
\citet{gl92b, gl96}.

\subsection{Fourier Analysis}
\label{sec:FFT}

We Fast Fourier Transformed the 420 million point time series
described in \S\ref{sec:prep} and searched the resulting Fourier
amplitudes using an advanced pulsar search code.  The search included
harmonic summing to improve sensitivity to low duty-cycle pulsations,
Fourier interpolation to minimize the effects of ``scalloping''
\citep{van89}, and the ability to compensate for signals with a
constant frequency derivative (i.e. an ``acceleration'' search) by
matched filtering of the complex Fourier amplitudes with a series of
template responses \citep{rgh+01, ran01}.

Due to the limitations in the HRC-S time resolution, we limited our
search to frequencies $f<100$\,Hz.  Similarly, we restricted the range
of acceptable frequency derivatives to be from $-5\times10^{-10} \le
\dot{f} \le 0$\,Hz\,s$^{-1}$, which encompasses the range observed for
all known isolated pulsars.  Assuming spin down via magnetic dipole
radiation, pulsars with period $P$ and magnetic field strengths $B >
3.2\times 10^{19}P^{3/2}T^{-1}$\,Gauss will drift across multiple
Fourier bins during observations of duration $T$ seconds.
Un-accelerated searches are considerably less sensitive to pulsars
with magnetic fields higher than this threshold.  For a 100\,ms period
and a 630\,ks time series (as for the DDT observation), this implies
reduced sensitivity for pulsars with $B > 1.6\times10^{12}$\,G,
thereby effectively eliminating a significant fraction of possible
phase space for such sources.  A near worst-case scenario for
un-accelerated searches is the Crab pulsar (for which $f = 29.82$\,Hz
and $\dot{f} = -3.7\times10^{-10}$\,Hz\,s$^{-1}$) which would drift by
$\left|\dot{f}T^2\right| \sim 149$ Fourier bins during the
observation.  In a raw power spectrum such a signal would be smeared
below detectability.  After accounting for the number of independent
trials searched, no candidates were detected with an equivalent
Gaussian significance of greater than 2\,$\sigma$.  For completeness,
we also searched for signals with positive frequency derivative from
$0\le \dot{f} \le 5\times10^{-10}$\,Hz\,s$^{-1}$ and detected no
significant pulsations.

In addition to searching the \emph{Chandra}\ DDT observation, we
searched the archival LETGS observation (ObsId 113, $\sim 55$\,ks) of
\ns\ as well as the 1997 October 9 \emph{ROSAT}\ High Resolution Imager
(HRI) observation ($\sim 29$\,ks) of \ns\ using virtually identical
techniques.  Not surprisingly, no candidates were found with
significance greater than 2\,$\sigma$ in either case.  In addition,
none of the lower significance candidates from these searches matched
any of the low-significance candidates from the DDT observation.

\subsection{Period Folding Analysis}
\label{sec:GLfold}

We conducted brute-force period folding of the time-filtered events
from each of the individual observations using a modified version of
the \citet{gl92b} technique that allows searching over $\dot{f}$ as
well as $f$.  This method maintains sensitivity to a wide variety of
pulsed signals by matching the complexity of a signal's pulse shape
with an appropriate number of bins in the folded profile aligned at
the optimal phase.  Successful use of this search algorithm has been
made by \citet{zps+00} and \citet{hhs+01}.

Each individual observation was searched using the time-filtered
events described in \S\ref{sec:prep} over a range of frequencies from
$0.001 < f <100$\,Hz and frequency derivatives from $-5\times10^{-10}
\le \dot{f} \le 0$\,Hz\,s$^{-1}$.  The candidate lists from each
observation were then compared to those from each of the other
observations to find likely matches in $f$ and $\dot{f}$.  No
interesting candidates were found.

Most of the CPU time used in the searches was spent searching the two
longest observations (ObsIds 3380 and 3381) together over a range of
frequencies $0.001 < f <500$\,Hz and frequency derivatives from
$-5\times10^{-10} \le \dot{f} \le 0$\,Hz\,s$^{-1}$.  The number of
operations required for a folding search that includes $\dot{f}$ goes
as $\sim N_{phot}T^2$, where $N_{phot}$ is the total number of photons
and $T$ is the total duration of the observation ($N_{phot} = 3582$
and $T = 395$\,ks for these two observations together).  This search
required over $10^{16}$ operations and produced no statistically
significant candidates.

\subsection{Pulsed Fraction Limits}
\label{sec:pfract}

Since no candidates were found in any part of our search, upper limits
on the pulsed fraction of \ns\ can be derived based on the predicted
response of a worst-case sinusoidal signal during the Fourier
analysis.  We use the standard definition of the pulsed fraction $f_p
= a / (a + b)$, where $a$ and $b$ are the pulsed and unpulsed count
rates respectively.  Since a sinusoidal signal in the presence of
noise with a total number of events $N_{phot} = (a+b)T$ produces an
expected normalized power\footnote{This normalization is a factor of
  two smaller than that used in some other X-ray timing papers.} of
$\left<P\right> = f_p^2 N_{phot}/4+1$, the pulsed fraction can be
written as $f_p =
\left[4\left(\left<P\right>-1\right)/N_{phot}\right]^{1/2}$ \citep[see
e.g.][]{van89,vvw+94}.

When no pulsation is detected during a search, an upper limit on the
pulsed fraction that the data could still contain can be calculated at
some level of confidence $C$, based on the maximum observed power in
the search.  \citet{vvw+94} describe how to calculate the signal power
that would be required to produce a measured power greater than the
maximum observed power a fraction $C$ of the time.  These calculations
are sensitive to how the data are binned and the number of independent
trials searched --- quantities that can be difficult to estimate.
Once this signal power has been calculated, substituting it for
$\left(\left<P\right>-1\right)$ in the definition for pulsed fraction
gives the upper limit on $f_p$ at a confidence level $C$.

Our best estimates for the limiting pulsed fraction for \ns\ based on
the new \emph{Chandra}\ observations, as well as the archival
\emph{Chandra}\ and \emph{ROSAT}\ observations, are given in
Table~\ref{table2}.  Slightly different upper limits are quoted based
on the fact that the number of trials is very different when comparing
an acceleration search with a more standard un-accelerated Fourier
analysis.  For the archival data sets, our 50\% confidence upper
limits for un-accelerated searches are roughly consistent with (but
slightly more conservative than) the values reported in \citet{bzn+01}
and \citet{pwl+02}.  For ``normal'' pulsars with relatively slow spin
periods (i.e. $f \la 5$\,Hz and $\dot{f} \la 10^{-12}$\,Hz\,s$^{-1}$),
the un-accelerated values are more appropriate.  

\section{Discussion}
\label{sec:discussion}

The 99\% confidence upper limits determined here of 4.1\% or 4.5\% for
un-accelerated or accelerated signals respectively, are the most
constraining limits yet on the pulsed fraction of X-ray emission from
\ns.  The fact that no pulsations have been observed in such sensitive
observations is a surprising result, especially considering the fact
that three other isolated NS candidates have measured pulsed fractions
of well over 10\% (see \S\ref{sec:intro}).  There are at least three
potential reasons for the lack of observed pulsations: 1) Radiation
from the NS is uniform in intensity and emitted isotropically
(possibly due to a uniform temperature distribution and a weak
magnetic field).  2) The viewing angle or emitting geometries are
unfavorable.  3) Gravitational bending effects near the NS decrease
the intrinsic pulsed fraction.  We will address each of these
possibilities in turn.
 
While uniform temperature blackbody models with $T\sim60$\,eV seem to
provide the best fits to the observed X-ray spectra, for the
$\sim60$\,pc distance of \citet{wal01} the implied NS ``radiation''
or ``redshifted'' radius $R_\infty = R/(1-2GM/Rc^2)^{1/2}$ is too
small ($R_\infty\sim2-4$\,km) for any known EOS \citep{bzn+01,
  pwl+02}.  While the new distance measurement of \citet{kva02}
increases the radiation radius to $R_\infty\sim5-9$\,km, since the
true radius is always less than the radiation radius these values are
still difficult to reconcile with most EOSs --- although self-bound
quark configurations (i.e. strange stars) may be possible
\citep{lp01}.  Unfortunately, an even bigger problem with the uniform
temperature blackbody models is that they under-predict the measured
optical flux by a factor of $\sim2$ \citep{pwl+02}.

If the surface temperature distribution of a NS is non-uniform, we may
expect to see X-ray pulsations at the star's rotation period resulting
from the motion of the hotter surface features (presumably polar cap
``hotspots'') through our line-of-sight.  There are at least two
geometries, though, that preclude the observation of such pulsations
no matter what the properties of the NS or its hotspots: when the
line-of-sight is aligned with the rotation axis of the star, or when
the polar cap axis is aligned with the rotation axis.  Neither of
these possibilities can be ruled out with the current observations,
and given the modulation strengths of the other observed isolated NSs,
this may be the simplest explanation for the lack of pulsations from
\ns.

Finally, gravitational bending effects on pulsations from hotspots on
neutron stars have been studied in detail by many groups
\citep*[e.g.][]{pfc83, rm88, zsp95, pod00}.  All have showed that the
measured pulsed fraction is a strong function of the angle between the
rotation axis and polar cap axis, the angle between the rotation axis
and the line-of-sight, the hotspot size as a fraction of the total
surface area ($\alpha$), and the mass and radius of the neutron star
itself.

For \ns, one can compensate for the under-prediction of the optical/UV
flux by the uniform temperature blackbody model (when fit to the X-ray
data alone) with the simple addition of a cool ($\sim20$\,eV)
blackbody component from the entire NS surface.  In this model, the
X-ray flux comes from thermal emission from much smaller (and
presumably polar cap) hotspots \citep{bzn+01, pwl+02}.  For the 60\,pc
distance to \ns, \citet{pwl+02} find acceptable two-component
blackbody fits to the optical, UV, and X-ray data with $\alpha \sim
0.2$, provided that $R_\infty \le 10$\,km and $M \le 1.3$\,\Msun\ (due
to the ``causality'' limit).  If such stars are possible, those with
``reasonable'' masses (i.e. $M\gtrsim1.2$\,\Msun) would be extremely
compact, with $R/M\lesssim 6$\,km\,$\Msun^{-1}$.  The extreme
gravitational fields of such compact NSs significantly bend the light
emitted from the surface and reduce the observed pulsed fraction.

\citet{pod00} calculated the likelihood of detecting pulsations of a
given pulsed fraction\footnote{The definition of pulsed fraction as
  given in eqn.~16 of \citet{pod00} is equivalent to that given in
  \S\ref{sec:pfract}.} after integrating over all possible polar cap
and line-of-sight angles as a function of $\alpha$ and the neutron
star radius-to-mass ratio.  Figure~6\footnote{Note that the axis of
  ordinates in this figure is mislabeled and should be a factor of ten
  lower.} of \citet{pod00} shows the fraction of neutron stars with
pulsed fraction greater than a specific value, plotted for several
values of the radius-to-mass ratio\footnote{Our definition of $p$,
  using the more traditional units km\,$\Msun^{-1}$ is equivalent to
  that of \citet{pod00}, who set $G=c=1$ giving $p \equiv Rc^2/2GM$.}
$p \sim 3 R/M$\,km\,$\Msun^{-1}$.  For stars as compact as those
implied by the 60\,pc two-component fit mentioned above, less than
$\sim20$\% yield pulsed fractions at or above the limits reported
here.

By transforming the two-component blackbody fit to 140\,pc and noting
that $\alpha$ is independent of the distance, we find $R_\infty \le
22$\,km.  This upper limit is compatible with virtually all modern
EOSs which predict $12 \lesssim R_\infty \lesssim 16$\,km
\citep{lp01}.  For a ``canonical'' neutron star with $R\sim11.5$\,km,
$M=1.35$\,\Msun, and $R/M \sim 3p \sim9$\,km\,$\Msun^{-1}$, the
majority ($\sim80$\%) would show pulsed fractions in excess of our
limits \citep{pod00}.  Similarly, except for highly unlikely and
unfavorable viewing geometries, stars with much higher values of $R/M$
(i.e. $\ga 12$\,km\,$\Msun^{-1}$) may be excluded by the lack of
pulsations from \ns.  These facts imply --- but do not require ---
that \ns\ is a relatively compact NS.

It is important to remember in this discussion of NS hotspots that no
truly reliable estimates of a NS radius to mass ratio can be derived
from blackbody spectral fits, since these fits only yield estimates of
the radiation radius ($R_\infty$) provided the distance is known.
Furthermore, the inferred area of the X-ray emitting region is
strongly dependent upon the presence and properties of any atmospheric
component.  Since there are as yet no atmospheric models consistent
with the current data \citep{bzn+01}, clearly, detailed spectral
analysis and modeling of the 2001-epoch \emph{Chandra} data should be
of great use in trying to explain the lack of pulsations from and the
overall nature of this enigmatic object.

\acknowledgements S.M.R. acknowledges the support of a Tomlinson
Fellowship awarded by McGill University.  B.M.G. acknowledges the
support of a Clay Fellowship awarded by the Harvard-Smithsonian Center
for Astrophysics.  P.O.S. acknowledges support from \emph{NASA}
contract NAS8-39073.  We would like to thank Steve Murray and Michael
Juda for providing additional information on time-filtering the HRC-S
events, and Dimitrios Psaltis and Feryal \"{O}zel for helpful
conversations on the effects of gravitational bending on pulsed
fraction.  Additional thanks go to Vicky Kaspi and the referee, Vadim
Burwitz, for careful readings of the manuscript.  This research has
made extensive use of \emph{NASA's} Astrophysics Data System
(\emph{ADS}) and High Energy Astrophysics Science Archive Research
Center (\emph{HEASARC}).  Many of the computations for this paper were
performed on equipment purchased with NSF grant PHY~9507695.


\begin{thebibliography}{29}
\expandafter\ifx\csname natexlab\endcsname\relax\def\natexlab#1{#1}\fi

\bibitem[{{Brinkman} {et~al.}(2000){Brinkman}, {Gunsing}, {Kaastra}, {van der
  Meer}, {Mewe}, {Paerels}, {Raassen}, {van Rooijen}, {Br{\" a}uninger},
  {Burkert}, {Burwitz}, {Hartner}, {Predehl}, {Ness}, {Schmitt}, {Drake},
  {Johnson}, {Juda}, {Kashyap}, {Murray}, {Pease}, {Ratzlaff}, \&
  {Wargelin}}]{bgk+00}
{Brinkman}, A.~C., {Gunsing}, C.~J.~T., {Kaastra}, J.~S., {van der Meer},
  R.~L.~J., {Mewe}, R., {Paerels}, F., {Raassen}, A.~J.~J., {van Rooijen},
  J.~J., {Br{\" a}uninger}, H., {Burkert}, W., {Burwitz}, V., {Hartner}, G.,
  {Predehl}, P., {Ness}, J.-U., {Schmitt}, J.~H.~M.~M., {Drake}, J.~J.,
  {Johnson}, O., {Juda}, M., {Kashyap}, V., {Murray}, S.~S., {Pease}, D.,
  {Ratzlaff}, P., \& {Wargelin}, B.~J. 2000, \apjl, 530, L111

\bibitem[{{Burwitz} {et~al.}(2001){Burwitz}, {Zavlin}, {Neuhaeuser}, {Predehl},
  {Truemper}, \& {Brinkman}}]{bzn+01}
{Burwitz}, V., {Zavlin}, V.~E., {Neuhaeuser}, R., {Predehl}, P., {Truemper},
  J., \& {Brinkman}, A.~C. 2001, \aap, 379, L35

\bibitem[{{Elvis} {et~al.}(1992){Elvis}, {Plummer}, {Schachter}, \&
  {Fabbiano}}]{eps+92}
{Elvis}, M., {Plummer}, D., {Schachter}, J., \& {Fabbiano}, G. 1992, \apjs, 80,
  257

\bibitem[{Gregory \& Loredo(1992)}]{gl92b}
Gregory, P.~C. \& Loredo, T.~J. 1992, \apj, 398, 146

\bibitem[{Gregory \& Loredo(1996)}]{gl96}
---. 1996, \apj, 473, 1059

\bibitem[{{Haberl} {et~al.}(1997){Haberl}, {Motch}, {Buckley}, {Zickgraf}, \&
  {Pietsch}}]{hmb+97}
{Haberl}, F., {Motch}, C., {Buckley}, D.~A.~H., {Zickgraf}, F.-J., \&
  {Pietsch}, W. 1997, \aap, 326, 662

\bibitem[{{Haberl} {et~al.}(1999){Haberl}, {Pietsch}, \& {Motch}}]{hpm99}
{Haberl}, F., {Pietsch}, W., \& {Motch}, C. 1999, \aap, 351, L53

\bibitem[{{Hambaryan} {et~al.}(2002){Hambaryan}, {Hasinger}, {Schwope}, \&
  {Schulz}}]{hhs+01}
{Hambaryan}, V., {Hasinger}, G., {Schwope}, A.~D., \& {Schulz}, N.~S. 2002,
  \aap, 381, 98

\bibitem[{{Kaplan} {et~al.}(2002){Kaplan}, {van Kerkwijk}, \&
  {Anderson}}]{kva02}
{Kaplan}, D.~L., {van Kerkwijk}, M.~H., \& {Anderson}, J. 2002, \apjl,
  submitted (astro-ph/0111174)

\bibitem[{{Lattimer} \& {Prakash}(2001)}]{lp01}
{Lattimer}, J.~M. \& {Prakash}, M. 2001, \apj, 550, 426

\bibitem[{Motch(2002)}]{mot02}
Motch, C. 2002, in AIP Conf. Proc. 599, X-ray Astronomy: Stellar Endpoints,
  AGN, and the Diffuse X-Ray Background, ed. N.~E. White, G.~Malaguti, \&
  G.~Palumbo (New York: AIP), 244

\bibitem[{{Murray} {et~al.}(1998){Murray}, {Chappell}, {Kenter}, {Kraft},
  {Meehan}, \& {Zombeck}}]{mck+98}
{Murray}, S.~S., {Chappell}, J.~H., {Kenter}, A.~T., {Kraft}, R.~P., {Meehan},
  G.~R., \& {Zombeck}, M.~V. 1998, in Proc. SPIE Vol. 3356, p. 974-984, Space
  Telescopes and Instruments V, Pierre Y. Bely; James B. Breckinridge; Eds.,
  Vol. 3356, 974--984

\bibitem[{{Pechenick} {et~al.}(1983){Pechenick}, {Ftaclas}, \& {Cohen}}]{pfc83}
{Pechenick}, K.~R., {Ftaclas}, C., \& {Cohen}, J.~M. 1983, \apj, 274, 846

\bibitem[{{Pons} {et~al.}(2002){Pons}, {Walter}, {Lattimer}, {Prakash},
  {Neuh{\" a}user}, \& {An}}]{pwl+02}
{Pons}, J.~A., {Walter}, F.~M., {Lattimer}, J.~M., {Prakash}, M., {Neuh{\"
  a}user}, R., \& {An}, P. 2002, \apj, 564, 981

\bibitem[{{Psaltis} {et~al.}(2000){Psaltis}, {{\" O}zel}, \& {DeDeo}}]{pod00}
{Psaltis}, D., {{\" O}zel}, F., \& {DeDeo}, S. 2000, \apj, 544, 390

\bibitem[{{Ransom}(2001)}]{ran01}
{Ransom}, S.~M. 2001, PhD thesis, Harvard University

\bibitem[{{Ransom} {et~al.}(2001){Ransom}, {Greenhill}, {Herrnstein},
  {Manchester}, {Camilo}, {Eikenberry}, \& {Lyne}}]{rgh+01}
{Ransom}, S.~M., {Greenhill}, L.~J., {Herrnstein}, J.~R., {Manchester}, R.~N.,
  {Camilo}, F., {Eikenberry}, S.~S., \& {Lyne}, A.~G. 2001, \apjl, 546, L25

\bibitem[{{Riffert} \& {Meszaros}(1988)}]{rm88}
{Riffert}, H. \& {Meszaros}, P. 1988, \apj, 325, 207

\bibitem[{{Tennant} {et~al.}(2001){Tennant}, {Becker}, {Juda}, {Elsner},
  {Kolodziejczak}, {Murray}, {O'Dell}, {Paerels}, {Swartz}, {Shibazaki}, \&
  {Weisskopf}}]{tbj+01}
{Tennant}, A.~F., {Becker}, W., {Juda}, M., {Elsner}, R.~F., {Kolodziejczak},
  J.~J., {Murray}, S.~S., {O'Dell}, S.~L., {Paerels}, F., {Swartz}, D.~A.,
  {Shibazaki}, N., \& {Weisskopf}, M.~C. 2001, \apjl, 554, L173

\bibitem[{{Treves} {et~al.}(2000){Treves}, {Popov}, {Colpi}, {Prokhorov}, \&
  {Turolla}}]{tpc+01}
{Treves}, A., {Popov}, S.~B., {Colpi}, M., {Prokhorov}, M.~E., \& {Turolla}, R.
  2000, in {Proc. of X-ray Astronomy 2000 (Palermo Sep. 2000), Eds. R.
  Giacconi, L. Stella, S. Serio; ASP Conf. Series}, in press (astro-ph/0011564)

\bibitem[{{van der Klis}(1989)}]{van89}
{van der Klis}, M. 1989, in Timing Neutron Stars, ({NATO ASI Series}), ed.
  H.~\"{O}gelman \& E.~P.~J. {van den Heuvel} (Dordrecht: Kluwer), 27--69

\bibitem[{{van Kerkwijk} \& {Kulkarni}(2001)}]{vk01}
{van Kerkwijk}, M.~H. \& {Kulkarni}, S.~R. 2001, \aap, submitted
  (astro-ph/0110065)

\bibitem[{{Vaughan} {et~al.}(1994){Vaughan}, {van der Klis}, {Wood}, {Norris},
  {Hertz}, {Michelson}, {van Paradijs}, {Lewin}, {Mitsuda}, \&
  {Penninx}}]{vvw+94}
{Vaughan}, B.~A., {van der Klis}, M., {Wood}, K.~S., {Norris}, J.~P., {Hertz},
  P., {Michelson}, P.~F., {van Paradijs}, J., {Lewin}, W. H.~G., {Mitsuda}, K.,
  \& {Penninx}, W. 1994, \apj, 435, 362

\bibitem[{{Walter}(2001)}]{wal01}
{Walter}, F.~M. 2001, \apj, 549, 433

\bibitem[{Walter \& Matthews(1997)}]{wm97}
Walter, F.~M. \& Matthews, L.~D. 1997, \nat, 389, 358

\bibitem[{{Walter} {et~al.}(1996){Walter}, {Wolk}, \& {Neuh{\" a}user}}]{wwn96}
{Walter}, F.~M., {Wolk}, S.~J., \& {Neuh{\" a}user}, R. 1996, \nat, 379, 233

\bibitem[{{Zavlin} {et~al.}(2000){Zavlin}, {Pavlov}, {Sanwal}, \& {Tr{\"
  u}mper}}]{zps+00}
{Zavlin}, V.~E., {Pavlov}, G.~G., {Sanwal}, D., \& {Tr{\" u}mper}, J. 2000,
  \apjl, 540, L25

\bibitem[{{Zavlin} {et~al.}(1995){Zavlin}, {Shibanov}, \& {Pavlov}}]{zsp95}
{Zavlin}, V.~E., {Shibanov}, Y.~A., \& {Pavlov}, G.~G. 1995, Astronomy Letters,
  21, 149

\end{thebibliography}

\clearpage

\begin{deluxetable}{ccccc}
  \footnotesize 
  \tablecaption{\emph{Chandra LETG/HRC-S} Observation
    Log} 
  \tablewidth{0pt} 
  \tablehead{\colhead{} &
    \colhead{$T_{\rm START}$} & \colhead{Exposure} &
    \colhead{Events} & \colhead{Events} \\
    \colhead{ObsId} & \colhead{(MJD)} & \colhead{(ks)} &
    \colhead{(Raw)\tablenotemark{a}} & 
    \colhead{(1\,ms Filter)\tablenotemark{b}}} 
  \startdata
  113  & 51613.33 & 55.5  & 10436 & 536  \\
  3382 & 52190.36 & 102.0 & 20072 & 1094 \\
  3380 & 52192.22 & 167.5 & 33703 & 1743 \\
  3381 & 52194.81 & 171.1 & 34571 & 1839 \\
  3399 & 52197.50 & 9.3   & 1788  & 104  \\
  \enddata 
  \tablenotetext{a}{These are the unfiltered events extracted
    from a $\sim2$\asec\ region around the nominal position of \ns:
    ${\rm RA (J}2000{ \rm)} = 18^{\rm h}$ 56$^{\rm m}$ 35\fs5,
    ${\rm DEC (J}2000{ \rm)} = -37\degrees\;54\amin\;36\farcs8$}
  \tablenotetext{b}{These events were time-filtered such that their
    arrival times differed by no more than 1\,ms from the event
    recorded after them (see \S\ref{sec:prep}).}
  \label{table1}
\end{deluxetable}

\begin{deluxetable}{cccccccccc}
  \footnotesize 
  \tablecaption{Pulsed Fraction Limits for \nsfull}
  \tablewidth{0pt} 
  \tablehead{\colhead{} &
    \multicolumn{4}{c}{No $\dot{f}$ Search} & \colhead{} &
    \multicolumn{4}{c}{With $\dot{f}$ Search} \\
    \cline{2-5} \cline{7-10} \\
    \colhead{ObsIds} & \colhead{$P_{max}$} & \colhead{$N_{trials}$} & 
    \colhead{$f_p^{50\%}$} & \colhead{$f_p^{99\%}$} 
    & \colhead{} & \colhead{$P_{max}$} & \colhead{$N_{trials}$}
    & \colhead{$f_p^{50\%}$} & \colhead{$f_p^{99\%}$}}

    \startdata
    400864\tablenotemark{a}         & 18.4 & 2.8 & $<$ 7.0\% & $<$ 9.8\% && 21.4 & 19 & $<$ 7.6\% & $<$ 10.3\% \\
    113\tablenotemark{b}            & 20.5 & 5.6 & $<$ 9.4\% & $<$ 12.8\% && 20.5 & 8.8 & $<$ 9.4\% & $<$ 12.8\% \\
    3380$-$2, 3399\tablenotemark{b} & 21.0 & 45  & $<$ 3.0\% & $<$ 4.1\% && 26.0 & 1340 & $<$ 3.4\%& $<$ 4.5\% \\
    \enddata 
    \tablenotetext{a}{\emph{ROSAT} HRI}
    \tablenotetext{b}{\emph{Chandra} LETGS}
    
    \tablecomments{These are the pulsed fraction upper limits for
      \nsfull\ based on the Fourier searches discussed in
      \S\ref{sec:FFT}.  The highest normalized power found in each
      search is denoted by $P_{max}$ and the approximate number of
      independent trials (including $\dot{f}$ trials and gaps in the
      data) is given by $N_{trials}$ in units of $10^6$\, trials.}
    
    \label{table2}
\end{deluxetable}

\end{document}